\documentclass{revtex4}
\usepackage{amssymb}
\usepackage{amsmath}

\input{tcilatex}

\begin{document}

\title{The Ordering Ambiguity}
\author{S. Habib Mazharimousavi}
\email{habib.mazhari@emu.edu.tr}
\affiliation{Department of Physics, Eastern Mediterranean University, G. Magusa, north
Cyprus, Mersin 10, Turkey.\\
Phon: +90-392-630-1067, Fax: +90-392-365-1604}
\pacs{PACS number}

\begin{abstract}
\textbf{Abstract:} The kinetic energy operator of a quantum particle with
position dependent mass and the associated ordering ambiguity is revisited.
We introduce a new form of this operator which is a continues or discreet
superposition of the acceptable values for the ordering parameters. 

\end{abstract}

\maketitle

\section{Introduction}

Position-dependent-mass (PDM) particles have received interests over the
last few decades \cite{1}. From the very beginning, parallel to the
applicability of the PDM-settings in different areas of physics such as the
many-body problem \cite{2}, semiconductors \cite{3}, quantum dots \cite{4},
quantum liquids \cite{5}, He-Clusters \cite{6}, etc. a mathematical
difficulty has grown which is known as \textit{the ordering ambiguity}.
Von-Roos has, firstly, reported the problem by proposing the revolutionary
form of the PDM kinetic energy operator which reads as \cite{7}%
\begin{equation}
\hat{T}=\frac{1}{4}\left( m^{\alpha }\hat{p}m^{\beta }\hat{p}m^{\gamma
}+m^{\gamma }\hat{p}m^{\beta }\hat{p}m^{\alpha }\right) .
\end{equation}%
Here $m$ is the position dependent mass of the physical particle and $\alpha
,$ $\beta $ and $\gamma $ are some real parameters which satisfy the
constraint $\alpha +\beta +\gamma =-1.$ In the constant mass setting, (1)
reduces to the standard kinetic energy operator which is given by 
\begin{equation}
\hat{T}=\frac{\hat{p}^{2}}{2m}.
\end{equation}%
We note that in Eq. (1), although $\alpha +\beta +\gamma =-1$ but the exact
values of $\alpha ,$ $\beta $ and $\gamma $ are not known and without any
harm to the mathematics of the problem, this von-Roos parameters may accept
any real number. The nature of the problem turns to the fact that the
momentum operator $\hat{p}$ does not commute with the position operator $%
\hat{x}$ in quantum theory.

In the literature, there exist several suggestions for the von-Roos ordering
parameters, for instance, the Gora's and Williams' ( $\beta =\gamma =0$, $%
\alpha =-1$) \cite{8}, Ben Daniel's and Duke's ($\alpha =\gamma =0$, $\beta
=-1$) \cite{9}, Zhu's and Kroemer's ( $\alpha =\gamma =-1/2$, $\beta =0$) 
\cite{10}, Li's and Kuhn's ($\alpha =0,$ $\beta =\gamma =-1/2$) \cite{11},
and the very recent Mustafa's and Mazharimousavi's ( $\alpha =\gamma =-1/4$, 
$\beta =-1/2$) \cite{12}. It has been observed that the physical
admissibility of a given ambiguity parameters set very well depends not only
on the continuity conditions at the abrupt heterojunction boundaries but
also on the position-dependent-mass form. The general consensus is that
there is no unique and universal choice for these ambiguity parameters.

\section{CONTINUES ORDERING PARAMETERS}

The idea which we shall expand in the sequel, is to construct a quantum
kinetic operator which is a superposition of all possible values for the
von-Roos parameters $\alpha ,$ $\beta $ and $\gamma .$ In doing so, we also
introduce a weight function which gives the distribution of the different
ordering parameters. The idea can be developed in two different directions:
i) a continues distribution and ii) a discrete distribution. Here in this
section we study the continues distribution and in the next section, the
discrete distribution will be given.

Following to our proposal we introduce the kinetic energy operator $\hat{T}$
for a position dependent mass particle as 
\begin{equation}
\hat{T}=\frac{1}{4A}\tiint\limits_{\mathcal{R}}dA\rho \left( \alpha ,\beta
\right) \left( m^{\alpha }\hat{p}m^{\beta }\hat{p}m^{\gamma }+m^{\gamma }%
\hat{p}m^{\beta }\hat{p}m^{\alpha }\right) 
\end{equation}%
in which $\rho \left( \alpha ,\beta \right) :%
\mathbb{R}
^{2}\rightarrow 
\mathbb{R}
$ is the weight function, $dA$ is the surface element on $\alpha -\beta $
plane and 
\begin{equation}
A=\tiint\limits_{\mathcal{R}}dA\rho \left( \alpha ,\beta \right) 
\end{equation}%
can be called a normalization constant (see Fig. 1 and 2). 

We note that the third parameter $\gamma $, is not free and therefore, the
integral is taken only over two of the parameters (i.e., $\alpha $ and $%
\beta $). Furthermore, one may impose additional conditions on these
parameters to specify the domain of the ordering parameters, for instance, $%
\alpha ,\beta ,\gamma \in \left[ -1,0\right] .$ Latter constraint looks to
be reasonable because, the well-known ordering-parameter sets available in
the literature, belong to this domain. In spite of this fact, we keep the
parameters free, to get any values in $%
\mathbb{R}
,$ but under a rather flexible constraint. 

Let's consider the domain of $\alpha ,$ $\beta $ and $\gamma $ the same and
fully symmetric which we called them $\mathcal{R}_{1},$ $\mathcal{R}_{2}$...
(Fig. 2). In this figure, $\mathcal{R}_{i}$ is located on the plane $\alpha
+\beta +\gamma =-1$ in the space of $\alpha ,$ $\beta $ and $\gamma .$ In
addition to $\mathcal{R}_{i}$ we also introduce $\mathcal{R}$ to be the
projection of $\mathcal{R}_{i}$ on the plane of $\alpha -\beta $. Finally,
we consider $\mathcal{R}_{i},$ an equilateral triangle which leads to 
\begin{equation}
\mathcal{R=}\left\{ 
\begin{array}{c}
-\left( 1+b\right) \leq \alpha \leq \frac{b}{2} \\ 
\\ 
-\left( 1+\frac{b}{2}+\alpha \right) \leq \beta \leq \frac{b}{2}%
\end{array}%
\right. 
\end{equation}%
in which $-\frac{2}{3}\leq b\in 
\mathbb{R}
$ and it shows the size of the region $\mathcal{R}_{i}$.

By considering the natural unit, i.e., $\hslash =1,$ one finds%
\begin{gather}
m^{\alpha }\hat{p}m^{\beta }\hat{p}m^{\gamma }+m^{\gamma }\hat{p}m^{\beta }%
\hat{p}m^{\alpha }= \\
-2\left( \left( 1+\alpha +\beta +\alpha \beta +\alpha ^{2}\right) \frac{%
m^{\prime 2}}{m^{3}}-\frac{1}{2}\left( 1+\beta \right) \frac{m^{\prime
\prime }}{m^{2}}-\frac{m^{\prime }}{m^{2}}\partial _{x}+\frac{1}{m}\partial
_{x}^{2}\right) ,  \notag
\end{gather}%
which yields%
\begin{equation}
\hat{T}=-\frac{1}{2A}\tiint\limits_{\mathcal{R}}dA\rho \left( \alpha ,\beta
\right) \left( \left( 1+\alpha +\beta +\alpha \beta +\alpha ^{2}\right) 
\frac{m^{\prime 2}}{m^{3}}-\frac{1}{2}\left( 1+\beta \right) \frac{m^{\prime
\prime }}{m^{2}}-\frac{m^{\prime }}{m^{2}}\partial _{x}+\frac{1}{m}\partial
_{x}^{2}\right) .
\end{equation}

\subsection{A UNIFORM DISTRIBUTION}

As an example, we study the case of $\rho \left( \alpha ,\beta \right) =1$
which is a uniform and symmetric distribution. By setting $\rho \left(
\alpha ,\beta \right) =1$ in Eq. (7) it leads to%
\begin{equation}
\hat{T}=\frac{-1}{6}\left( \frac{1}{16}\left( 3b^{2}+4b+28\right) \frac{%
m^{\prime 2}}{m^{3}}-\frac{m^{\prime \prime }}{m^{2}}\right) +\frac{1}{2}%
\left( \frac{m^{\prime }}{m^{2}}\partial _{x}-\frac{1}{m}\partial
_{x}^{2}\right) .
\end{equation}%
It is remarkable to observe that for $\forall b\in \left[ -\frac{2}{3}%
,\infty \right) $ non of the first two terms are zero. We note that among
the possible values for $b,$ $b=-\frac{2}{3}$ corresponds to 
\begin{equation}
\alpha =\beta =\gamma =-\frac{1}{3}.
\end{equation}%
This set of parameters consequently implies 
\begin{equation}
\hat{T}=\frac{-1}{2}\left( \frac{5}{9}\frac{m^{\prime 2}}{m^{3}}-\frac{1}{3}%
\frac{m^{\prime \prime }}{m^{2}}\right) +\frac{1}{2}\left( \frac{m^{\prime }%
}{m^{2}}\partial _{x}-\frac{1}{m}\partial _{x}^{2}\right) .
\end{equation}%
Another interesting case is when $b=0$, which makes $\alpha ,$ $\beta $ and $%
\gamma $ lie in the interval $\left[ -1,0\right] .$ The kinetic energy after
setting $b=0$ reads%
\begin{equation}
\hat{T}=\frac{-1}{6}\left( \frac{7}{4}\frac{m^{\prime 2}}{m^{3}}-\frac{%
m^{\prime \prime }}{m^{2}}\right) +\frac{1}{2}\left( \frac{m^{\prime }}{m^{2}%
}\partial _{x}-\frac{1}{m}\partial _{x}^{2}\right) .
\end{equation}

\subsection{A NON-UNIFORM DISTRIBUTION}

Next we give an example with the non-uniform distribution function $\rho
\left( \alpha ,\beta \right) $ which is given by%
\begin{equation}
\rho \left( \alpha ,\beta \right) =\frac{1}{\left( \alpha ^{2}+1\right) }+%
\frac{1}{\left( \beta ^{2}+1\right) }.
\end{equation}%
This is a symmetric distribution with respect to $\alpha $ and $\beta .$
Consequently the form of kinetic energy becomes%
\begin{equation}
\hat{T}=-\frac{1}{2}\left( \eta _{1}\frac{m^{\prime 2}}{m^{3}}+\eta _{2}%
\frac{m^{\prime \prime }}{m^{2}}-\frac{m^{\prime }}{m^{2}}\partial _{x}+%
\frac{1}{m}\partial _{x}^{2}\right) ,
\end{equation}%
where%
\begin{equation}
\eta _{1}=\frac{-\left( 3b^{2}+12b+20\right) \ln \left( \frac{4\left(
2+2b+b^{2}\right) }{4+b^{2}}\right) +2\left( b^{3}+3b^{2}+12b+4\right)
\left( \tan ^{-1}\frac{b}{2}+\tan ^{-1}\left( 1+b\right) \right)
+15b^{2}+40b+20}{48\left( 1+b\right) \left( \tan ^{-1}\frac{b}{2}+\tan
^{-1}\left( 1+b\right) \right) -24\ln \left( \frac{4\left( 2+2b+b^{2}\right) 
}{4+b^{2}}\right) },
\end{equation}%
and%
\begin{equation}
\eta _{2}=\frac{\left( b+4\right) \ln \left( \frac{4\left( 2+2b+b^{2}\right) 
}{4+b^{2}}\right) +2\left( 3b+2\right) \left( \tan ^{-1}\frac{b}{2}+\tan
^{-1}\left( 1+b\right) \right) -\left( 3b+2\right) }{16\left( 1+b\right)
\left( \tan ^{-1}\frac{b}{2}+\tan ^{-1}\left( 1+b\right) \right) -8\ln
\left( \frac{4\left( 2+2b+b^{2}\right) }{4+b^{2}}\right) }
\end{equation}%
One can check that 
\begin{equation}
\lim_{b\rightarrow \left( -\frac{2}{3}\right) ^{+}}\eta _{1}=\frac{5}{9},
\end{equation}%
and%
\begin{equation}
\lim_{b\rightarrow \left( -\frac{2}{3}\right) ^{+}}\eta _{2}=-\frac{1}{3}
\end{equation}%
which are exactly as we expected for $\alpha =\beta =\gamma =-\frac{1}{3}.$
Also for the case of $b=0$ one gets%
\begin{equation}
\eta _{1}=\frac{\pi +10\left( 1-\ln 2\right) }{6\left( \pi -2\ln 2\right) }%
=0.589\,66,\text{ \ \ }\eta _{2}=-\frac{\pi +2-4\ln 2}{4\left( \pi -2\ln
2\right) }=-0.337\,46.
\end{equation}

\section{Discrete ordering parameters}

As we mentioned before, there are some well-known ordering-parameters sets 
\cite{8,9,10,11,12} which have been used widely in the literatures. These
different sets which have been introduced for various physical problems,
could be an indication of having a discrete ordering parameters. In this
line let's introduce the kinetic energy operator with discreet distribution
as%
\begin{equation}
\hat{T}=\frac{\sum\limits_{\alpha ,\beta }C_{\alpha ,\beta }\left( m^{\alpha
}\hat{p}m^{\beta }\hat{p}m^{\gamma }+m^{\gamma }\hat{p}m^{\beta }\hat{p}%
m^{\alpha }\right) }{4\sum\limits_{\alpha ,\beta }C_{\alpha ,\beta }},
\end{equation}%
in which $C_{\alpha ,\beta }$ is a real number which can be called
distribution number. We again recall that $\gamma =-1-\alpha -\beta $ and in
the summation $\gamma $ is not a free index. Very similar to the case of
continues distribution one may rewrite%
\begin{equation}
\hat{T}=-\frac{1}{2}\frac{\sum\limits_{\alpha ,\beta }C_{\alpha ,\beta }%
\left[ \left( 1+\alpha +\beta +\alpha \beta +\alpha ^{2}\right) \frac{%
m^{\prime 2}}{m^{3}}-\frac{1}{2}\left( 1+\beta \right) \frac{m^{\prime
\prime }}{m^{2}}-\frac{m^{\prime }}{m^{2}}\partial _{x}+\frac{1}{m}\partial
_{x}^{2}\right] }{\sum\limits_{\alpha ,\beta }C_{\alpha ,\beta }}.
\end{equation}%
To see how this may work, we consider an equal possibility for all known
ordering given before, i.e., i) Gora and Williams ( $\beta =\gamma =0$, $%
\alpha =-1$), ii) Ben Danial and Duke ($\alpha =\gamma =0$, $\beta =-1$),
iii) Zhu and Kroemer ($\alpha =\gamma =-1/2$, $\beta =0$), iv) Li and Kuhn ($%
\alpha =0,$ $\beta =\gamma =-1/2$) and v) Mustafa and Mazharimousavi ($%
\alpha =-1/4,$ $\beta =-1/2=-1/4$). Considering these values into (20) and
setting $C_{\alpha ,\beta }=1,$ we find 
\begin{equation}
\sum\limits_{\alpha ,\beta }C_{\alpha ,\beta }=5
\end{equation}%
and consequently%
\begin{equation}
\hat{T}=-\frac{1}{10}\left( \frac{43}{16}\frac{m^{\prime 2}}{m^{3}}-\frac{3}{%
2}\frac{m^{\prime \prime }}{m^{2}}\right) +\frac{1}{2}\left( \frac{m^{\prime
}}{m^{2}}\partial _{x}-\frac{1}{m}\partial _{x}^{2}\right) .
\end{equation}

\section{CONCLUSION}

In this Letter, we have considered the long standing ordering ambiguity
problem associated to the von-Roos kinetic energy operator for PDM
particles. A superposition of the possible values of the von-Roos parameters
has been used to construct a general kinetic energy operator. We have also
introduced the distribution function which can be continues or discreet. We
believe that this kinetic energy operator can assist the theoretical
physicists to adjust their results along the line of the relevant
experiments. In the other words there would be more free parameters in the
theoretical calculations.

\textbf{Figure Captions:}

Fig. 1: The plane of $\alpha +\beta +\gamma =-1$ in the space of $\alpha
\beta \gamma .$ The surface element on this plane is given too.

Fig. 2: The possible symmetric regions of $\alpha ,$ $\beta $ and $\gamma $
on the plane of $\alpha +\beta +\gamma =-1$. The projection of these regions
on $\alpha \beta -$plane is give by (5).

\end{document}